\documentclass[3p,times]{elsarticle}

\usepackage{ecrc}

\volume{00}

\firstpage{1}

\journalname{Pervasive and Mobile Computing}

\runauth{}

\jnltitlelogo{Pervasive and Mobile Computing}

\usepackage{amssymb}

\usepackage[figuresright]{rotating}

\usepackage{amssymb,amsmath}
\usepackage{graphicx}
\usepackage{subfigure}

\usepackage{float}
\restylefloat{table}

\begin{document}

\begin{frontmatter}



\dochead{}

\title{Scalable Oriented-Service Architecture for Heterogeneous and Ubiquitous IoT Domains}


\author{
Pablo Lopez, 
David Fernandez, 
Rafael Marin-Perez, 
Antonio J. Jara, 
Antonio F. Gomez-Skarmeta, 
}



\address{Department of Information and Communications Engineering, Computer Science Faculty, University of Murcia, Regional Campus of International Excellence "Campus Mare Nostrum", Murcia, Spain}

\begin{abstract}
Internet of Things (IoT) grows quickly, and 50 billion of IoT devices will be interconnected by 2020. 
For the huge number of IoT devices, a high scalable discovery architecture is required to provide autonomous registration and look-up of IoT resources and services.
The architecture should enable dynamic updates when new IoT devices are incorporated into Internet, and changes are made to the existing ones. 
Nowadays in Internet, the most used discovery architecture is the Domain Name System (DNS).
DNS offers a scalable solution through two distributed mechanisms: multicast DNS (mDNS) and DNS Service Directory (DNS-SD).
Both mechanisms have been applied to discover resources and services in local IoT domains. 
However, a full architecture has not still been designed to support global discovery, local directories and a search engine for ubiquitous IoT domains. 
Moreover, the architecture should provide other transversal functionalities such as a common semantic for describing services and resources, and a service layer for interconnecting with M2M platforms and mobile clients.
This paper presents an oriented-service architecture based on DNS to support a global discovery, local directories and a distributed search engine to enable a scalable looking-up of IoT resources and services.
The architecture provides two lightweight discovery mechanisms based on mDNS and DNS-SD that have been optimized for the constraints of IoT devices to allow autonomous registration.
Moreover, we analyse and provide other relevant elements such semantic description and communications interfaces to support the heterogeneity of IoT devices and clients.
All these elements contribute to build a scalable architecture for the discovery and access of heterogeneous and ubiquitous IoT domains.

\end{abstract}

\begin{keyword}
IoT, M2M, IPv6, architecture, service, repository.
\end{keyword}

\end{frontmatter}



\section{Introduction}
\label{Introduction}
Internet grows quickly in the number of devices and users interconnected. 
The wide range of interconnection will be addressed by the Internet Protocol version 6 (IPv6).
IPv6 is the standard proposed for the addressing and networking of a globally connected Internet of Things~\cite{0p}.
The Internet of Things will offer an extra value to real-world applications such as Smart Cities and Building Automation.
IoT devices should be discoverable, accessible, available, usable, and interoperable through IPv6 technologies.
The emergence of IPv6-related standards specifically designed for the IoT, such as 6LoWPAN and CoAP~\cite{14p,15p}, has enabled highly constrained devices, also called smart objects, to become natively IP compliant. 

Recently, several projects have designed different IoT architectures depending on their specific applications and requirements (SENSEI ~\cite{6p}, HOBNET ~\cite{7p}, iCORE~\cite{29p}, BUTLER ~\cite{30p}, FI-WARE~\cite{10p}, IoT-A~\cite{8p}, etc.). 
Due to a large heterogeneity of applications and requirements, the architecture approaches differ between the projects resulting in more or less different components and protocols.
However, IERC (Internet of Things European Research Cluster) recognizes that the architectures diversity is the main factor limiting the advance of the IoT technology.

In addition, the fast evolution of the IoT technology is defining new architecture challenges in term of scalability, allocation of resources and efficient discovery. 
It needs to be determined among the extended set of existing mechanisms for resource and service discovery which are appropriated for the IoT requirements. 
For that reason, we analyse the major relevant requirements to achieve an oriented-service architecture for IoT domains:
\begin{itemize}
  \item Scalability: 
It is estimated that over 50 billion devices will be connected to Internet by 2020 ~\cite{53d}. 
This implies that a high amount of resources and services will need to be managed. 
Therefore, a decentralized architecture is required such as DNS which distributes the information about the services deployed by Internet devices.
Thereby, the information can be managed locally, but be accessible globally through Internet.
  \item Dynamic: 
Smart objects are being deployed continuously; therefore new devices and services will be continually defined. 
In addition, some smart objects will be mobile (i.e. Intelligent Transport Systems).
Therefore, the architecture should support automatically the creation, update and delete of the registration about services available from smart objects. 
  \item Communication constraints: 
IoT technologies such as IEEE 802.15.4 have a small frame size of 127 bytes. 6LoWPAN protocol has an overload of 26-41 bytes, this means that the final available payload is reduced to the size half (61-76 bytes).
Therefore, it will require a high fragmentation of payload into multiple frames increasing highly the overload of IoT devices constrained in terms of bandwidth and energy.

  \item Global query capability:  
A resource directory must be provided to register IoT resource and services and support queries over them at local domain.
The architecture must provide a distributed search engine to look-up globally the domains where the queried resources or services are available.

  \item Semantic description: 
A common description of the services is needed to carry out the queries. It is a collateral requirement to define the mechanisms to filter adequately the type of resources and services to be queried.

  \item Heterogeneity of things and clients: 
Many heterogeneous things can connected to Internet through different technologies such as IPv6, 6LoWPAN or legacy (e.i. CAN, X10, EIB/KNX and BACNet). Moreover, the architecture must support the most relevant client applications (i.e. DNS and HTTP) to access Internet resources.

  \item Based on existing Internet technologies:
The architecture should be based on existing mechanisms in Internet. 
But these mechanisms must be adapted to reach a trade-off among all the presented requirements.
Therefore, the architecture could be based on DNS and its extensions (mDNS and DNS-SD) or could be built over the application level (i.e. HTTP). 

\end{itemize}

These requirements are difficult to be satisfied by existing IoT architecture. 
These requirements often are interrelated, even some requirements influence directly to another ones. 
For instance, the semantic description defines the common format for global queries.

According to the defined requirements, this paper presents an oriented-service architecture to support global and scalable discovery of ubiquitous resources and services offered by heterogeneous IoT devices.
The proposed architecture provides global and local discovery based on a core system and distributed repositories at local domains following the DNS infrastructure.
The core system called \textit{digcovery} allows looking-up globally IoT resources and services according to their types, domains and locations.
Each distributed repository called \textit{digrectory} registers the resource and services provided by IoT devices in each local domain. The architecture enables the publishing of resources and services in the \textit{digrectories} into the core \textit{digcovery} system.
Moreover, this scalable architecture provides a search engine to discover and access ubiquitous resources and services from all the \textit{digrectories} based on DNS-SD in ZeroConf IETF Working Group.
Moreover, the architecture provides two optimizations of the mDNS protocol to minimize the payload fragmentation in constrained IoT devices\cite{a0}.
Moreover, we analyse and provide other relevant elements such as semantic description and communication interfaces to support heterogeneous IoT devices and client applications.
All these contributions enable a scalable oriented-service architecture for discovering, registering and looking-up heterogeneous and ubiquitous IoT resources and services.

The remainder of the paper is organized as follows. 
Section 2 describes the related work about existing protocols, technologies and architectures for IoT domains.
Section 3 presents the oriented-service architecture proposed for the IoT requirements.
Section 4 provides concluding remarks and future work.

\section{Related Work}
This section is organized following the evolution of the main IoT challenges: interconnection of IoT devices, building applications over them and defining techniques to discover resources and services.
Initial work has been performed to offer IPv6 connectivity to IoT devices based on IEEE 802.15.4, Bluetooth Low Energy, BACNET, etc. This work is contextualized mainly under the 6LoWPAN ~\cite{2d}, GLoWBAL IPv6 ~\cite{3d}, IPv6 addressing Proxy~\cite{0b} and 6man ~\cite{4d, 5d} for the IPv6 integration.

Once the end-to-end connection for any IoT device is available through Internet, an homogeneous access to the application layer must be defined. 
Analysing the current Internet status, the Web is the most extended services medium and therefore, the Web of Things ~\cite{6d} was defined, in which, at the beginning, RESTFul packets (HTTP) were carried over 6LoWPAN. This solution was seen as highly flexible and powerful. Moreover, a reduced version of RESTFul for constrained devices was proposed as COAP (Constrained Application Protocol). 

Once we have access to IoT resources and services, an scalable solution to discover them is needed.
Two discovery levels are found according to an Internet of Things and Ubiquitous computing approach~\cite{7d}.
The first level is resource discovery, i.e. the discovery of devices on the network.
And the second level is the service discovery, the discovery of the services, methods and functions offered by a specific resource. 
Usually, Internet is considered as a resource from a general point of view where services are part of this resource. 
However, when Internet is not limited to just files, applications, and services, and is moved towards a more physical approach, the physical location and identification of the information is also required.
Resources are reachable through technologies such as 6LoWPAN, GLoWBAL IPv6 or any technology offering IPv6 support. 
Resource discovery is the process by which the user is able to find devices offering services according to his criteria and interests. It can differ from the resources that the user can explicitly request or from a more sophisticated discovery where the network is more pro-active, and it notifies the user about the availability of new devices. 
Resource discovery will provide descriptive information, such as the resource type or family, and some attributes to describe it. 
In addition, resource discovery will provide the information that the user will need to reach them. 
This information will be a locator such as URL, UID, Host Identity (HIP) or IP address.
Resource discovery requires dynamic updates when new resources are included in the network, as well as the ability to integrate the updates over mobile devices~\cite{8d}.
Service discovery is focused on the description of those services provided by technologies such as Web Services.
These services include printing and file transfer, music sharing, servers for pictures. 
Simpler services can be considered with the expansion towards the Internet of Things such as the pressure value for a parking sensor.

Different techniques can be found for resources and services discovery in IoT domains. 
Currently, the most common approach is the definition of M2M platforms, such as ThingWorx, Pachube, Sen.Se, and SENSEI, where the devices are registered in the platform, and are reachable through Web Services such as SOAP and REST. 
The problem with this static approach is that it is limited to the information on the platform and the manually registered devices. Therefore, a more scalable solution is required to enable autonomous registration of new IoT resouces without any human interaction.
To do that, local directory is needed for homogeneous and simple queries over Internet without the requirement of using a specific M2M platform.

Existing solutions that require a manual and static management of resources are considered infeasible for IoT domains.
Some naming systems such as Lightweight Directory Access Protocol (LDAP), Universal Description, Discovery, and Integration (UDDI), and Domain Name System (DNS) ~\cite{9d} offer resource and service directory capabilities.
In particular, more flexible resource discovery technologies based on UPnP, JINI, Service Location Protocol (SLP), and Rendezvous or Bonjour protocols over DNS~\cite{10d} could be used.
However, none of the existing implementations are considered the constraints of IoT devices in terms of computing, battery, memory or bandwidth. 
Considering the IoT constraints and requirements, we design a service-oriented architecture providing scalable and autonomous discovery mechanisms, as shown in next Section.

\section{Scalable Oriented-Service Architecture for Heterogeneous and Ubiquitous IoT Domains}

\begin{figure*}[htbp]
\begin{center}
\includegraphics[width=0.5\textwidth]{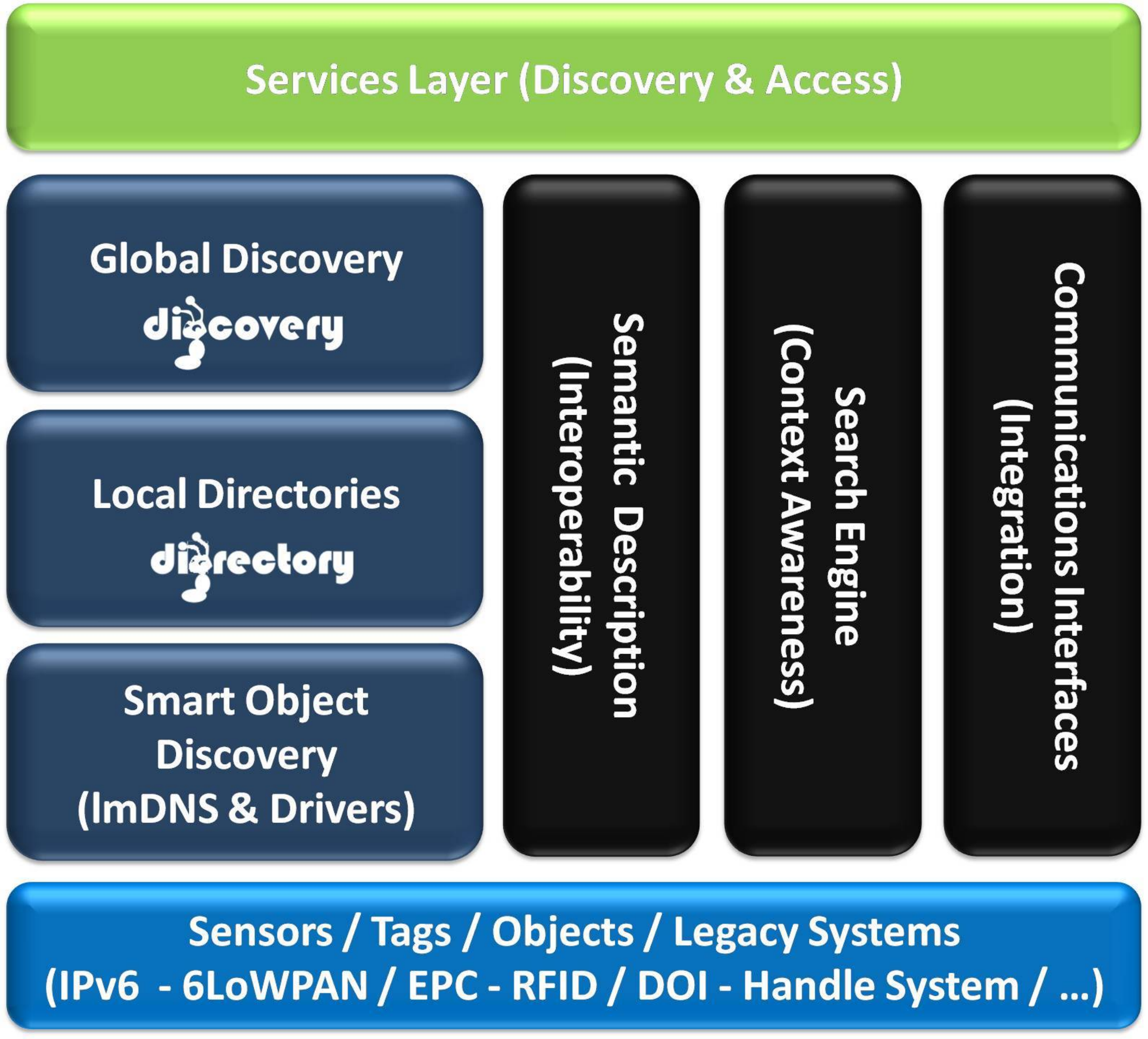}  
\caption{Key components of IoT architecture} 
\label{components}
\end{center}
\end{figure*}

This section presents a service-oriented architecture for supporting the global discovery and homogeneous access to IoT devices.
The architecture considers the necessary functionalities providing an unifying framework over the heterogeneity and fragmentation of the IoT. 
The considered functionalities are global discovery, local directories, resource discovery, search engine, semantic description and communication interfaces.
In the proposed architecture, these functionalities are covered by the components shown in Figure~\ref{components}.

The top green component represents a service layer to support the interaction with end-user applications through standard interfaces.
These interfaces can be naming applications (i.e. DNS), web services (i.e. HTTP), or constrained applications (i.e. COAP).

The left dark blue components are the main elements proposed to enable a scalable architecture to discover, look-up and register IoT services and resources. 
\textit{Digcovery} is the core platform that enables the looking-up of services in local domains managed by \textit{digrectories}. 
Each \textit{digrectory} contains the descriptions of the resources and services in a local domain. 
Moreover, a smart object discovery protocol based on existing discovery mechanisms is included to enable the interaction between IPv6-enabled devices and \textit{digrectories}. 
Specifically, the smart object protocol provides a lightweight version of multicast DNS (mDNS) and Service Discovery (DNS-SD) to discover services and resources of IPv6-enabled devices. 
These proposed elements are presented in detail in Section~\ref{sec:digcovery-digrectory}.

The right black components have been mainly analysed and adapted to satisfy the needs of the proposed elements (dark blue ones).
First, a semantic description is very important to support homogeneous looking-ups of resources and services from heterogeneous IoT networks. 
Section~\ref{sec:semantic} presents the existing semantic approaches such as IPSO Alliance which is chosen for the proposed architecture.
Second, a search engine is key for any scalable discovery solution.
In particular, ElasticSearch with some extensions is integrated to support context awareness look-ups based on geo-location, domain and type of resources (see Section~\ref{sec:search}).
Third, communication interfaces are needed to support the interoperability the proposed elements with heterogeneous IoT things and clients. 
The interoperability is guaranteed through existing interfaces based on standardized technologies such as IPv6, DNS, RESTFul and COAP as shown in Section~\ref{sec:communication-interfaces}.

\begin{figure*}[htbp]
\begin{center}
\includegraphics[width=0.5\textwidth]{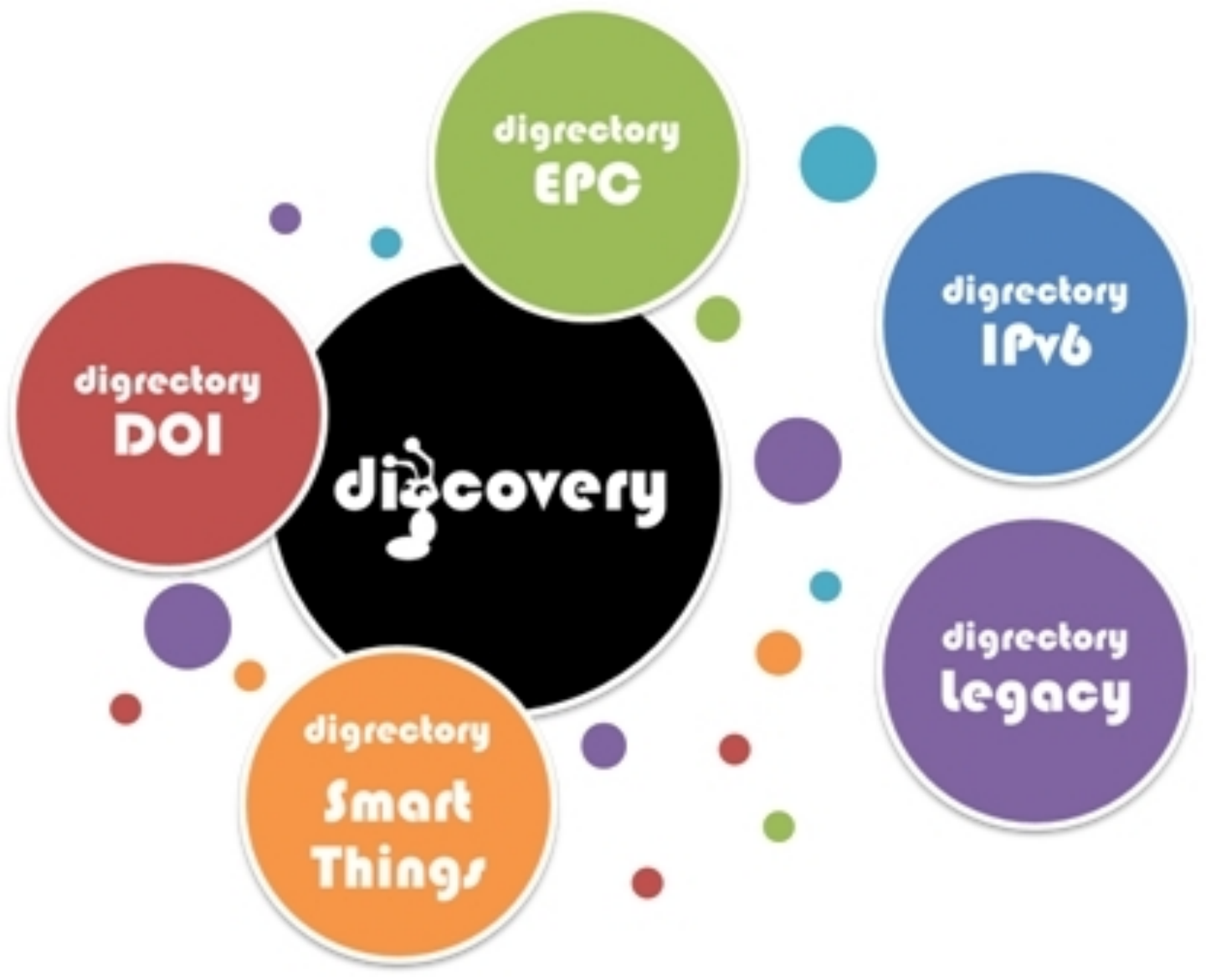}  
\caption{Resources ecosystem} 
\label{ecosystem}
\end{center}
\end{figure*}

The proposed service-oriented architecture considers the wide heterogeneity of IoT devices. 
The Internet of Things ecosystem is composed not only of IPv6-enabled devices with communications and processing capabilities, but also of non-IP devices based on legacy technologies such as RFID, BACnet, etc.
Legacy devices have been developed to satisfy the requirements of specific applications as follows. 
RFID and NFC were designed for transportation in terms of logistics and ticketing, respectively. 
Healthcare devices are used for continuous patient monitoring through wireless sensor networks.
BACnet was developed for Building Automation with proprietary protocols for lighting, heating, cooling, and security. 
Smart meters and smart grid applications require mainly reliability. 
Smart tags are replacing barcodes which require energy-efficiency, low-cost, real-time, and scalability.
The aforementioned requirements have led to different technical alternatives and standards from ETSI such as M2M architectures focused on cellular networks.
These provide support for reliability, they have low energy-efficiency, high cost and relatively poor scalability. 
IETF with the Working Groups (i.e. ROLL, 6LoWPAN, CORE, LWIG, and COMA) is focused on constrained devices and sensors to provide low cost protocols with high energy-efficiency and scalability, but with no mobility support.
EPCGlobal for RFID provides low cost, real-time wireless identification, but the sensing capabilities are not standardized yet.
Finally, the Handle System for Digital Object Identifiers offers low cost, real-time, identification and does not rely on a pre-defined medium to carry out the identification.
For these reasons, the design of the proposed architecture considers the heterogeneity of the IoT resources as shown Figure~\ref{ecosystem}.
The IoT resources are mainly smart objects identified through IPv6 addressing. 
Smart objects are connected through technologies such as 6LoWPAN, GLoWBAL IPv6 and lwIP. 
Moreover, other non-IP devices have been considered such as Digital Object Identifier (DOI) via the Handle System, RFID resources identified through Electronic Product Code (EPC) and legacy technologies(i.e. BACnet, X10, etc).
The integration of these heterogeneous IoT devices is handled by \textit{digrectories} at local domain level.
In addition, local \textit{digrectories} offer all their services and resources to the global \textit{digcovery} system, as described in the next section.

\subsection{Global Digcovery, Local Digrectories and Smart Object Discovery}
\label{sec:digcovery-digrectory}
This section presents three main components proposed to offer a scalable oriented-service architecture for heterogeneous IoT subsystems.
The location of each component is described below.
First, global \textit{digcovery} is a core service system located in a cloud platform.
Second, local \textit{digrectory} is a resource repository deployed as a middleware distributed in routers or gateways of IoT networks.
Third, smart object discovery is placed in IoT devices supporting native IPv6 communication based on 6lowpan, Glowbal-IPv6, etc.

The proposed architecture is based on the combination of two main approaches: centralized and decentralized. 
A centralized approach depends on a well-known server storing all the services.
A decentralized approach depends on several directories saving the services in local domains. 
Both approaches present a set of advantages and disadvantages. 
Therefore, we propose a combined approach which collects the advantages of centralized approach in terms of easily accessible and well-known access point, and also the advantages from decentralized approach in terms of scalability.

Using the combined approach, the central \textit{digcovery} platform offers a full vision of all available resources in the distributed \textit{digrectories}. 
\textit{Digcovery} and \textit{digrectories} manage the services and resources provided by ubiquitous IoT devices.
Each \textit{digrectory} saves information detailed about the services and resources of IoT devices in the local domain.
The \textit{digcovery} stores information simplified about types of services available in the \textit{digrectories}. 
This combined approach for managing services and resources at global and local levels will be described in Section~\ref{sec:digcovery-mechanisms}.
Moreover, the architecture employs a search engine and a common semantic description to support global queries for heterogeneous IoT devices, as explained in Sections~\ref{sec:semantic} and ~\ref{sec:search}, respectively.

In addition, smart object discovery allows updating automatically the available services when new IoT devices are incorporated into the network.
The smart object discovery is also responsible of replying to the information queries from the local \textit{digrectory}. 
Therefore, this presents a double role as client of configuration and server of resources. 
In the case of non-IP legacy technologies (i.e. BACnet, RFID, etc), these functionalities are delegated to the \textit{digrectories} with specific drivers to adapt the proprietary protocols to a common interface based on DNS.
The smart object discovery is based on a lightweight version of the mDNS protocol detailed in the next Section. 

\subsection{Smart Object Discovery}   
\label{sec:smart-objects-discovery-protocol}

A smart object discovery is proposed based on the DNS protocol, which is the main technique for discovery services in Internet. 
For self-discovery, IPv6 devices employ the mDNS protocol based on multicast messages to publish their services at the network level.
Also, a local \textit{digrectory} uses mDNS to request information from services in a local network.
mDNS is carried out by commercial solutions such as Bonjour in Apple products, and Avahi for Linux-based platforms. 
Moreover in mDNS messages, it employs the DNS-SD format for indicating the available services.
This section presents optimizations and recommendations to make lightweight approaches of mDNS and DNS-SD for reducing control overhead in constrained IoT devices~\cite{0a}.
First, we introduce the most common records from DNS:
\begin{itemize}
  \item A: Address record for an IPv4 address.
  \item AAAA: Address record for an IPv6 address.
  \item CNAME: Alias from one name to another name.
  \item NS: Delegation of a DNS Zone to an authoritative name server.
  \item Others: MX for the email, HIP for the HIP identifier, LOC for the locater and others related with security stuff. 
\end{itemize}

A smart object discovery should start with a multicast DNS message (mDNS). 
The discovery of these services can be carried out from a mDNS client such as Avahi, Bonjour or the DNS look-up tool (e.g. Dig command).
For a specific type of service, a local \textit{digrectory} can send a mDNS message to discover the IoT devices offering the service required at local domain. 
For this purpose, the PTR record defines multiples pointers to a IoT device depending on its family, type, services, etc. 
For example, Table~\ref{tab3} presents multiple PTRs pointing to a light from our lab, called $light\_lab$.
\begin{table*}[htbp]
\begin{displaymath}
\begin{array}{|l|}
\hline
;Type \\
\_lamp.\_sub.\_coap.\_udp\ PTR\ light\_lab \\
;Services \\
\_status.\_lamp.\_sub.\_coap.\_udp\ PTR\ light\_lab \\
\_onoff.\_lamp.\_sub.\_coap.\_udp\ PTR\ light\_lab \\
\_dimmer.\_lamp.\_sub.\_coap.\_udp\ PTR\ light\_lab \\
;Technology \\
\_x10.\_lamp.\_sub.\_coap.\_udp\ PTR\ light\_lab \\
\hline
\end{array}
\end{displaymath}
\caption{PTR record for the service light\_lab} 
\label{tab3}
\end{table*}

\begin{itemize}
  \item PTR: is used for the reverse DNS look-ups, i.e. from address to name. 
  But, this presents a total different usage for mDNS and DNS-SD. 
  PTR is used to filter the queries in mDNS and to describe the services in DNS-SD.
\end{itemize}
In addition, mDNS and DNS-SD provide extra functionalities for the records SRV and TXT. Below, we describe both records with optimizations to reduce their sizes.

\begin{table*}[htbp]
\begin{displaymath}
\begin{array}{|l|}
\hline
;light\_lab.rd.esiot.com\ SRV \\
;;\ ->>HEADER<<-\ opcode:\ QUERY,\ status:\ NOERROR,\ id:\ 6373 \\
;;\ flags:\ qr\ rd\ ra;\ QUERY:\ 1,\ ANSWER:\ 1,\ AUTHORITY:\ 0,\ ADDITIONAL:\ 0 \\
\\
;;\ QUESTION\ SECTION: \\
;light\_lab.rd.esiot.com.\ \ \ IN\ SRV \\
;;\ ANSWER\ SECTION: \\
light\_lab.rd.esiot.com.\ 604800\ IN\ SRV\ 0\ 0\ 1234\ light1.rd.esiot.com. \\
;;\ Query\ time:\ 118\ msec \\
;;\ MSG\ SIZE\ rcvd:\ 79 \\
\hline
\end{array}
\end{displaymath}
\caption{SRV query of the found light with optimizations} 
\label{tab5}
\end{table*}

Once PTR records provide the hostname of a searching service, the SRV record is used to request the service.
Table~\ref{tab5} shows a SRV record that defines the $light$ service located at the hostname $light\_lab.rd.esiot.com$. 
\begin{itemize}
  \item SRV is a generalized service location record similar to MX but for any service. 
SRV describes which machine supports what service and on what port. 
The syntax is: $SRV [priority] [capacity] [ttl] [hostname]$.
Priority and capacity parameters allow choosing a hostname among different devices when they are offering the same service. 
\end{itemize}

\begin{table*}[htbp]
\begin{displaymath}
\begin{array}{|l|}
\hline
;\ light1.rd.esiot.com\ TXT \\
;;\ Got\ answer: \\
;;\ ->>HEADER<<-\ opcode:\ QUERY,\ status:\ NOERROR,\ id:\ 19187 \\
;;\ flags:\ qr\ rd\ ra;\ QUERY:\ 1,\ ANSWER:\ 1,\ AUTHORITY:\ 0,\ ADDITIONAL:\ 0 \\
;;\ QUESTION\ SECTION: \\
;light1.rd.esiot.com.\ \ \ \ IN\ TXT \\
;;\ ANSWER\ SECTION: \\
light1.rd.esiot.com.\ 604800\ IN\ TX\ "rt=light\;ins=2\;lt=86400\;model=dimmer\; \\
if=EIB\;area=1\;zone=2\;deviceID=3\;value\;onoff" \\
;;\ Query\ time:\ 79\ msec \\
;;\ MSG\ SIZE\ rcvd:\ 130 \\
\hline
\end{array}
\end{displaymath}
\caption{TXT query of the found light with optimizations for in a single TXT record} 
\label{tab7}
\end{table*}

One a device with the required service is chosen, the TXT entry is employed to obtain detail information about the device.
\begin{itemize}
  \item TXT: This contains extra information (metadata) for the device. 
  The metadata format is $'[key]':'[value]'$, and the contents depend on the protocol. 
\end{itemize}

TXT entries are designed to be associated with the hostname of the SRV entry. 
Usually, each TXT entry is defined as a single '[key]':'[value]' per record. 
To reduce the number of records, they can be associated by services, resource type (rt) and interface (if).
In addition, TXT entries can be joined in an unique record as shown in Table~\ref{tab7}.
This presented format follows the naming conventions that describe how services will be represented in DNS records, 
as defined by Web Linking format under the CoRE IETF working group~\cite{12d}.

When all the information about the hostname of the resource (SRV) and service description with extra information (TXT) have been obtained, the IP address of the device is required. 
AAAA records provide IPv6 address of devices reachable through technologies such as 6LoWPAN and GLoWBAL IPv6.
Moreover, $A$ records can be considered for backward compatibility with the current Internet infrastructure based on IPv4.
Also, these records enable other addressing and identification spaces such as Universal Identifier (UID) from RFID or novel protocols such as Host Identity Protocol (HIP) as described in~\cite{0b}.

\begin{table*}[htbp]
\begin{displaymath}
\begin{array}{|l|}
\hline
;;\ search(light\_lab.rd.esiot.com,\ SRV,\ IN) \\
;;\ query(light\_lab.rd.esiot.com,\ SRV,\ IN) \\
\\
;;\ send\_udp(94.142.247.17:53):\ sending\ 40\ bytes \\
;;\ timeout\ set\ to\ 5\ seconds \\
;;\ answer\ from\ 94.142.247.17:53:\ 188\ bytes \\
;;\ HEADER\ SECTION \\
;;\ id\ =\ 9950 \\
;;\ qr\ =\ 1\ opcode\ =\ QUERY\ aa\ =\ 0\ tc\ =\ 0\ \ \ rd\ =\ 1 \\
;;\ ra\ =\ 1\ \ \ rcode\ =\ NOERROR \\
;;\ qdcount\ =\ 1\ ancount\ =\ 1\ nscount\ =\ 1\ arcount\ =\ 4 \\
\\
;;\ QUESTION\ SECTION\ (1\ record) \\
;light\_lab.rd.esiot.com.\ IN\ SRV \\
;;\ ANSWER\ SECTION\ (1\ record) \\
light\_lab.rd.esiot.com.\ 602400\ IN\ SRV\ 0\ 0\ 1234\ light1.rd.esiot.com. \\
\\
;;\ AUTHORITY\ SECTION\ (1\ record) \\
rd.esiot.com.\ \ 602400\ IN\ NS\ rd.esiot.com. \\
\\
;;\ ADDITIONAL\ SECTION\ (4\ records) \\
light1.rd.esiot.com.\ 602512\ IN\ A\ 155.54.210.163 \\
light1.rd.esiot.com.\ 602400\ IN\ AAAA\ 2001:720:1710::11 \\
rd.esiot.com.\ \ 602400\ IN\ A\ 155.54.210.159 \\
rd.esiot.com.\ \ 602400\ IN\ AAAA\ 2001:720:1710:0:216:3eff:fe00:9 \\
\hline
\end{array}
\end{displaymath}
\caption{SRV query of the found light without optimizations} 
\label{tab9}
\end{table*}

In conclusion, Table~\ref{tab9} presents an equivalent query to that carried out in Table~\ref{tab5} for the discovery of SRV entries associated to the service to consult. 
the SRV original information presents a packet size of 188 bytes instead of 79 bytes. 
This means that a packet with the optimizations fits in a single frame, while the original use of DNS-SD and mDNS requires 3 frames.
Moreover, the description of the services (TXT) should be simplified as much as possible in order to fill in a single frame. 
Therefore, the TXT have been joined in an unique entry following some formats such as the aforementioned link format~\cite{12d}. 
The difference can be seen between the original query in Table~\ref{tab10} and the optimized version in Table~\ref{tab7} which is 130 bytes instead of 221 bytes. 
Also, the TXT entry should be simplified and reduced further to values under 80 bytes making it feasible for a single 6lowpan packet of 127 bytes. 
This reduction could come through the use of wildcards for the identification of the parameter types, or through compression techniques such as LZ77.

\begin{table*}[htbp]
\begin{displaymath}
\begin{array}{|l|}
\hline
;;\ search(light\_lab.rd.esiot.com,\ TXT,\ IN) \\
;;\ query(light\_lab.rd.esiot.com,\ TXT,\ IN) \\
;;\ send\_udp(94.142.247.17:53):\ sending\ 40\ bytes \\
;;\ timeout\ set\ to\ 5\ seconds \\
;;\ answer\ from\ 94.142.247.17:53:\ 221\ bytes \\
;;\ HEADER\ SECTION \\
;;\ id\ =\ 24910 \\
;;\ qr\ =\ 1\ \ \ opcode\ =\ QUERY\ \ aa\ =\ 0\ \ tc\ =\ 0\ rd\ =\ 1 \\
;;\ ra\ =\ 1\ \ \ rcode\ =\ NOERROR \\
;;\ qdcount\ =\ 1\ ancount\ =\ 3\ nscount\ =\ 1\ arcount\ =\ 2 \\
\\
;;\ QUESTION\ SECTION\ (1\ record) \\
;light\_lab.rd.esiot.com.\ IN\ TXT \\
\\
;;\ ANSWER\ SECTION\ (3\ records) \\
light\_lab.rd.esiot.com.\ 604800\ IN\ TXT\ "if=X10;housecode=A;unitcode=5" \\
light\_lab.rd.esiot.com.\ 604800\ IN\ TXT\ "rt=light;ins=1;lt=86400;model=normal" \\
light\_lab.rd.esiot.com.\ 604800\ IN\ TXT\ "onoff;status;dimmer" \\
;;\ AUTHORITY\ SECTION\ (1\ record) \\
rd.esiot.com.\ \ 604800\ IN\ NS\ rd.esiot.com. \\
;;\ ADDITIONAL\ SECTION\ (2\ records) \\
rd.esiot.com.\ \ 604800\ IN\ A\ 155.54.210.159 \\
rd.esiot.com.\ \ 604800\ IN\ AAAA\ 2001:720:1710:0:216:3eff:fe00:9 \\
\hline
\end{array}
\end{displaymath}
\caption{TXT query of the found light without optimizations} 
\label{tab10}
\end{table*}

\subsection{Global Discovery and Local Directory}
\label{sec:digcovery-mechanisms}

Global discovery requires the management of different domains within a single core system.
We propose a core system called \textit{digcovery}, since it is based on DNS ($dig$ command in Linux OS and MAC OS system).
\textit{Digcovery} is public and accessible from any place through a Web portal $www.digcovery.net$. 
In addition, the \textit{digcovery} platform is accessed via standardized technologies such as DNS and CoAP, as shown Section~\ref{sec:communication-interfaces}.
\textit{Digcovery} allows the delegation of each domain to the end-user and/or service providers through local \textit{digrectories}.
The proposed architecture enables the publishing and linking of IoT resources and services registered in local \textit{digrectories} to the global \textit{digcovery}.
Moreover, \textit{digcovery} is really a cloud-based platform, therefore, it is highly scalable around the world.

Local directory is performed by \textit{digrectories} interacting directly with the IoT devices at network level.
Each \textit{digrectory} provides a detailed enumeration of all resources and services in a local domain. 
\textit{Digrectories} are located in gateways or routers and employ the DNS-SD protocol to store the detailed information of services from IoT devices. 
DNS-SD is a fine-grained service description and enables the concept of discovering services according to their properties. 
DNS-SD supports a hierarchical approach to the naming of services and allows a decentralized \textit{digrectory} infrastructure that scales well with the network size.

In the following, we describe the global and local discovery solution that is divided into three main phases: registration, discovery and resolution.
The registration phase manages services and resources of IoT devices. 
\textit{Digrectories} employ the format defined by DNS-SD (i.e. PTR, SRV and TXT) to store the services and resources available in local domain.
Each \textit{digrectory} control the adding, updating and removing through specific protocols such as mDNS for IPv6 devices, lmDNS for smart object, etc.
The discovery phase lists all the services under some queried criteria. 
A search engine is used by the \textit{digcovery} platform to find the queried services, as described in Section~\ref{sec:search}.
The search engine is built over the PTR records. Note that the instances of the SRV records are PTR records. This query is mainly based on subtypes (e.g. \_light.\_coap.udp). 
Once the appropriated service is chosen (i.e. PTR for DNS), the resolution phase requests detailed information at the local domain.
The resolution query is send directly to the \textit{digrectory} containing the chosen service.
The \textit{digrectory} responds DNS-SD information such as service description (SRV), resources/attributes description (TXT), and device addressing (A/AAAA).
Moreover, \textit{digrectories} publish the PTR pointers of available services into the \textit{digcovery} system.
The resolution phase is divided into two subphases: look-up and query.
Look-up subphase gets the information of a service instance: service name, hostname, port, etc (e.i. SRV and TXT). 
Query subphase obtains the IP address of a resource (hostname) from records in DNS-SD (e.i. A and AAAA).

The following subsections present these three phases for IPv6 smart things and non-IP legacy technologies.
First, we describe how to register an smart things through the proposed lightweight mDNS protocol.
Second, we explain how to discover services and resources of smart things in local domains.
Third, we show how to discover special resources of legacy technologies(i.e. EPCIS tags of RFID) integrated through \textit{digrectories}.

\begin{figure*}[htbp]
\begin{center}
\includegraphics[width=0.8\textwidth]{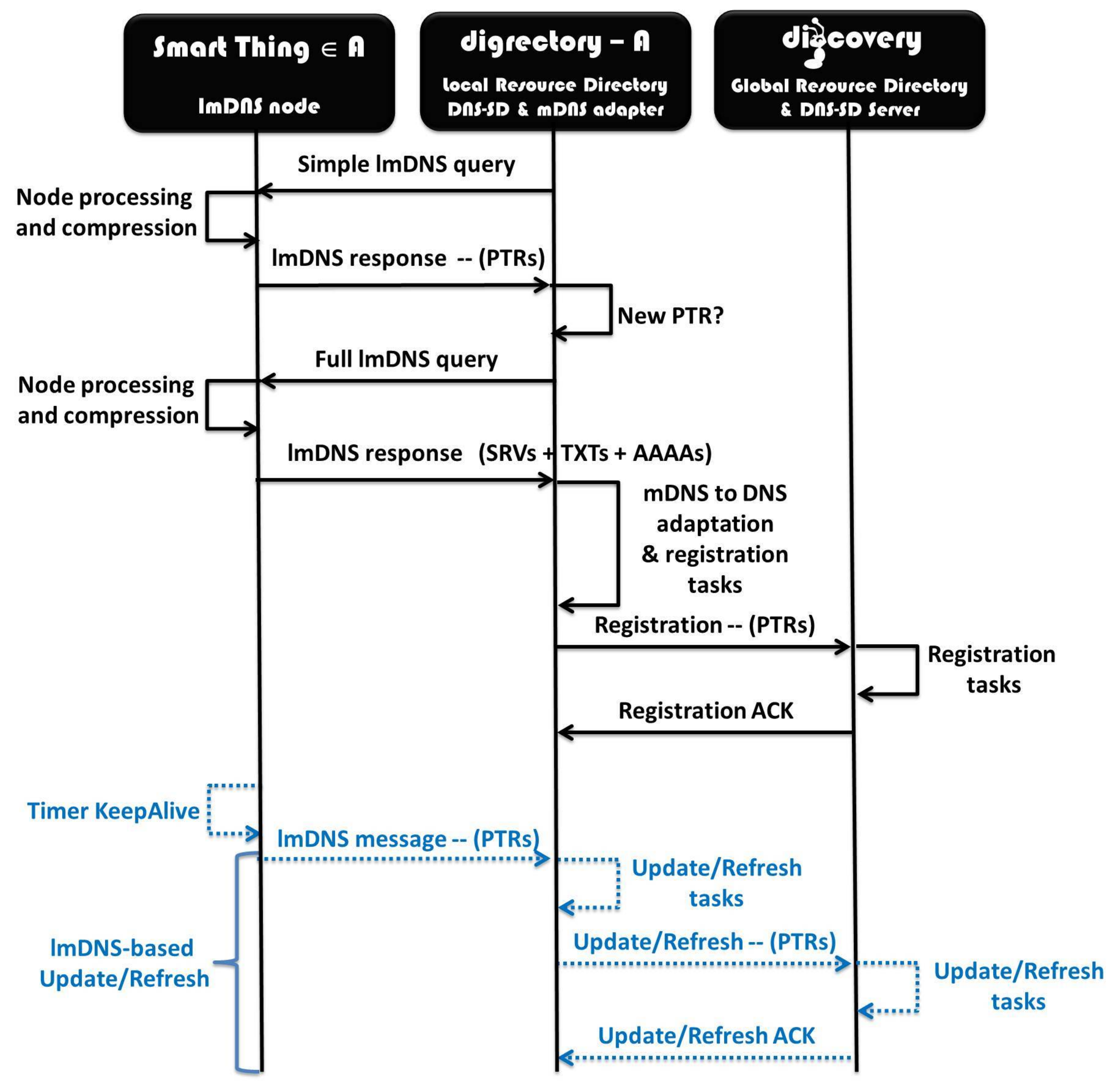}  
\caption{Registration of IPv6-based things using the lmDNS proposal} 
\label{registration}
\end{center}
\end{figure*}

\subsubsection{Registration of IPv6-based Things}

The registration procedure for IPv6-based things is shown in Figure~\ref{registration} by a sequence diagram.
This registration is based on the proposed lightweight version of multicast DNS (lmDNS) described in Section~\ref{sec:smart-objects-discovery-protocol}.
The diagram presents the initial phase of the registration process, which is triggered by a lmDNS query from the local \textit{digrectory}.
Only if a new PTR record exists, the resolution phase requests all services information such as service description (SRV), resources/attributes descriptions(TXT), and device addressing (A/AAAA). 
The main goal is to reduce the power consumption from the IPv6-based things.
In addition to the registration process, the IPv6-based things will be refreshed periodically regarding its entry into the \textit{digrectory}, to ensure the freshness and integrity of the \textit{digrectory} information.

\subsubsection{Discovery and Resolution of IPv6-based Things}

\begin{figure*}[htbp]
\begin{center}
\includegraphics[width=0.9\textwidth]{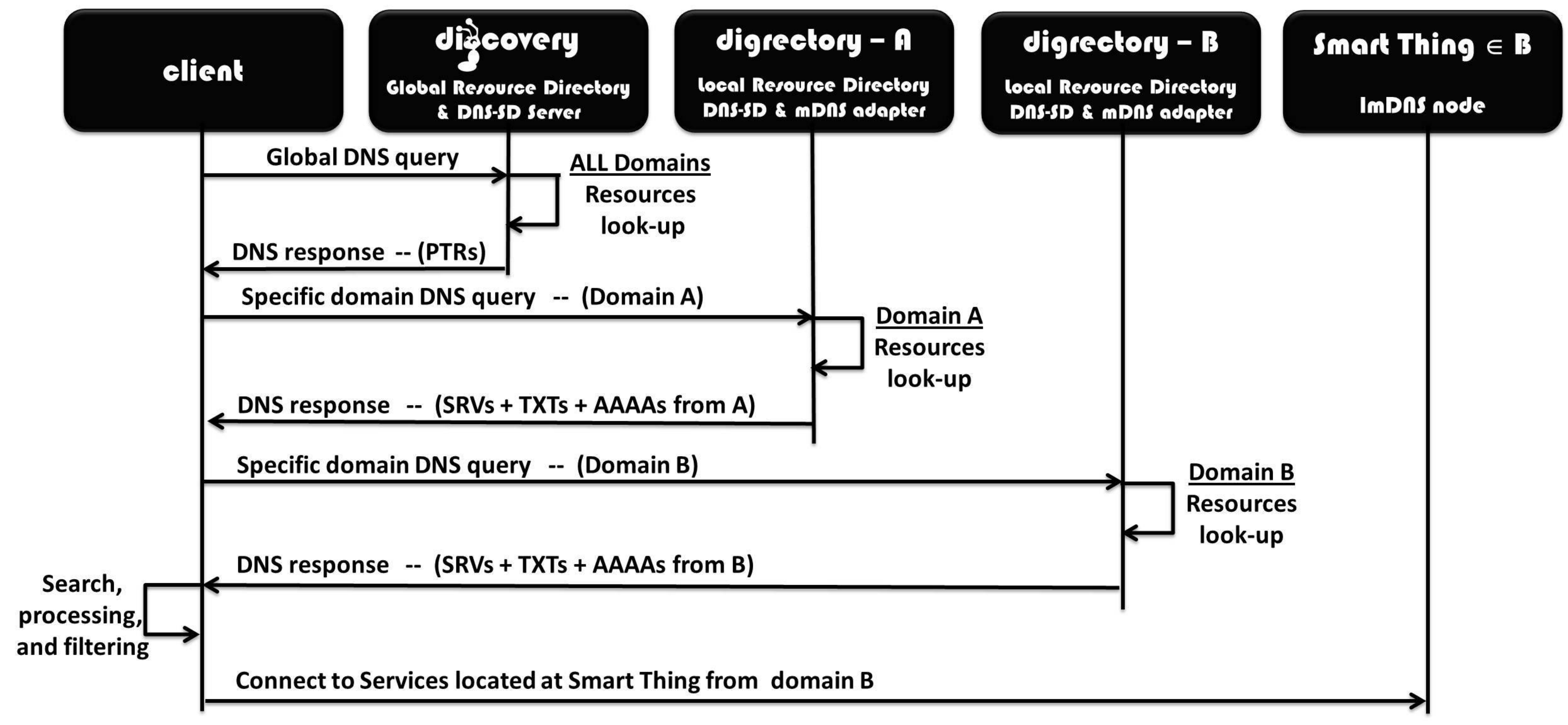}  
\caption{IPv6-based resource and service discovery} 
\label{discovery-phase}
\end{center}
\end{figure*}

Discovery and resolution procedures are shown in the sequence diagram of Figure~\ref{discovery-phase}.
The sequence begins when the \textit{digcovery} receives a global DNS query from the client. 
This queries can be expressed with regular expressions (i.e. *.*). 
Then, \textit{digcovery} looks-up in all the domains to find those IoT things with the queried services/resources. 
To do that, \textit{digcovery} employ a search engine to distribute global queries to all \textit{digrectories}, as described in Section~\ref{sec:search}.
Using the search engine, \textit{Digcovery} obtains the list of domains where the services/resources are found and sends this list to the client.
The user client requests directly through DNS to those \textit{digrectories} with the queried services/resources. 
These \textit{digrectories} provide detailed information about the queried services/resources (i.e. TXT records with the extended information, SRV record with the service description, and AAAA record with the IPv6 address). 

\subsubsection{Discovery and Resolution of Non-IP Things}

In addition, the architecture considers more complex queries for the integration of legacy technologies with non-IP things. 
An example of this extended query is related with the integration of RFID tags. 
Specifically, the RFID integration is carried out through the EPC Information System (EPCIS). 
For that reason, it is required to integrate query mechanisms from the DNS system to the EPCIS.

\begin{figure*}[htbp]
\begin{center}
\includegraphics[width=1\textwidth]{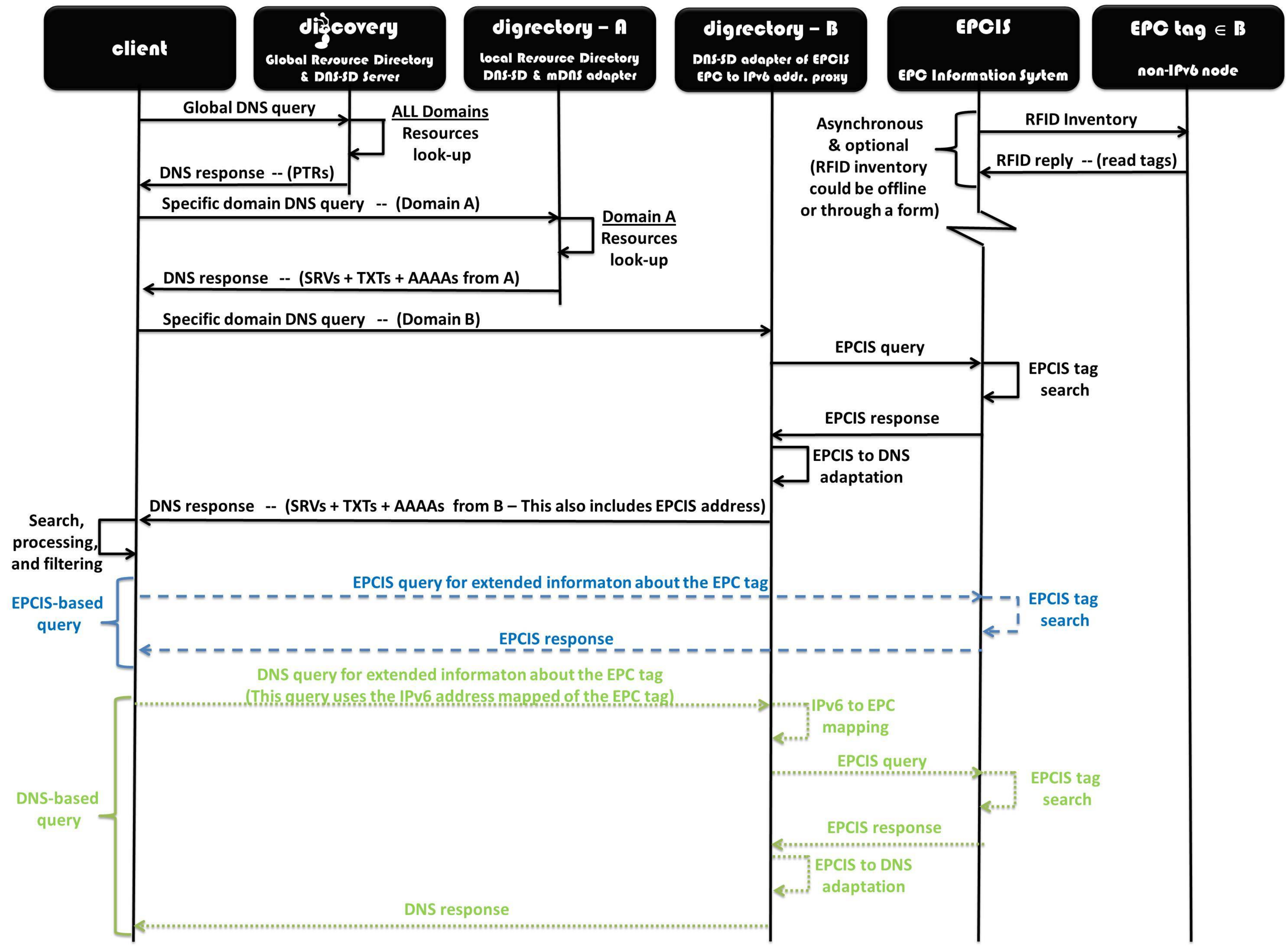}  
\caption{Non-IP resource and service discovery} 
\label{discovery-phase-nonIP}
\end{center}
\end{figure*}

The proposed solution for non-IP devices is presented in Figure~\ref{discovery-phase-nonIP}.
In the domain $B$, a local \textit{digrectory} is connected directly to the EPCIS system. 
The \textit{digrectory B} provides an adaptation between EPCIS API and DNS queries.
In addition, the \textit{digrectory B} provides the mapping between EPC identifier and IPv6 address, as we described in~\cite{0b}.
Both adaptations enable the exchange of queries and responses between the EPCIS server and the \textit{digrectory}.
Once the client discover a RFID tag, there are two ways to collect its attributes, features and extended information. 
First, the client may request directly to the EPCIS server using its proprietary API (represented in blue colour). 
Second, the client may query to the \textit{digrectory} employing the DNS protocol (represented in green colour).

\subsection{Semantic Services Description}
\label{sec:semantic}
To support the common queries for heterogeneous IoT domains, a homogeneous description of the services and attributes must be defined. 
This is a collateral requirement to define the mechanisms to query and filter adequately the type of resources and services. 
Currently, several different ways are being developed to establish a common representation for queries in IoT domains. Some of the work in this field is identified below.
Specifically, the IPSO Alliance is defining a common family of interfaces and resource types for Resource Directory from CoRE ~\cite{60d}. 
This could be re-used in a similar way as it is re-used in the link format. 
IPSO is defining on the one hand, a simple set of interfaces based on CoAP and plain text and on the other hand, a more structured version based on JSON with the semantic from SenML.
As an alternative, more complex solutions such as Triple Spaces on RDF ~\cite{61d} can be defined which allows the retrieval, creation, modification or deletion of resources in the RDF graphs. 
This knowledge representation is based on a common ontology, which all the entities involved in the communication sharing. 
The queries over RDF can follow a pattern similar to what was defined in CoAP based on a triple pattern with wildcards (e.g. the format for queries) or also more sophisticated and complex solutions such as SPARQL. 
In the following, we introduce each semantic description defined for the Internet of Things.
Moreover, we will provide a comparative analysis among these semantic solutions and indicate the chosen one for the proposed architecture.

\subsubsection{IPSO Alliance Interfaces (IETF)}
Constrained RESTful Environments (CoRE) working group specifies a set of CORE interfaces for IoT constrained nodes and networks~\cite{1b}. 
The IETF draft defines several functionalities that cover the needs of the IoT transmission technologies. 
The defined functionalities consist of a set of mandatory inputs, outputs and parameters to provide the minimum M2M interoperability. 
The IETF draft proposes a common representation for the binding between M2M devices that specifies a format based on the CoRE link format document~\cite{11d}. 
This format represents the binding information accompanied by a set of rules to define a binding method as a specialized relationship between two M2M resources.
As defined in the CoRE Resource Directory~\cite{55d}, all resources and services offered by a device should be discoverable either through a direct link in $/.well-known/core$ or by following successive links starting from $/.well-known/core$. Table~\ref{tab11} illustrates the discovery procedure, as defined the IETF draft.

\begin{table*}[htbp]
\begin{displaymath}
\begin{array}{|l|}
\hline
Req:\ GET\ /.well-known/core \\
Res:\ 2.05\ Content (application/link-format) \\
</s>;rt="simple.sen";if="core.b", \\
</s/lt>;rt="simple.sen.lt";if="core.s", \\
</s/tmp>;rt="simple.sen.tmp";if="core.s";obs, \\
</s/hum>;rt="simple.sen.hum";if="core.s", \\
</a>;rt="simple.act";if="core.b", \\
</a/1/led>;rt="simple.act.led";if="core.a", \\
</a/2/led>;rt="simple.act.led";if="core.a", \\
</d>;rt="simple.dev";if="core.ll",  \\
</l>;if="core.lb",  \\
\hline
\end{array}
\end{displaymath}
\caption{Example of service discovery~\cite{1d}} 
\label{tab11}
\end{table*}

The draft describes CORE interfaces for Link List, Batch, Sensor, Parameters, Actuators and Binding. 
Some variants such as Linked Batch or Read-Only Parameter are also defined. 
The interfaces support the usage of plain text and/or SenML Media types to specify the payload. 
Table~\ref{tab12} shows a relation of methods defined for each resource where the column “if=” indicates the Interface Description attribute value, as used in the CoRE Link Format.

\begin{table*}[htbp]
\begin{displaymath}
\begin{array}{|l|c|c|}
\hline
\textbf{Interface} & \textbf{if=} & \textbf{Methods}  \\
\hline
Link List	  & core.ll	& GET \\
\hline
Batch	      & core.b	& GET, PUT, POST (where applicable) \\
\hline
Linked Batch & core.lb & GET, PUT, POST, DELETE (where applicable) \\
\hline
Sensor	    & core.s & GET \\
\hline
Parameter	  & core.p & GET, PUT \\
\hline
Read Only Parameter	& core.rp	& GET \\
\hline
Actuator	  & core.a & GET, PUT, POST \\
\hline
Binding	    & core.bnd & GET, POST, DELETE \\
\hline
\end{array}
\end{displaymath}
\caption{Defined interfaces in the IETF draft~\cite{1d}} 
\label{tab12}
\end{table*}

First, the Link List interface retrieves (GET) a list of resources where the request should contain an Accept option with the application/link-format content type. 
This option may be elided if the resource does not support any other form. 
This request returns a list of URI references expressed as an absolute path to the resources.
Second, the Batch interface manipulates a collection of sub-resources.
This interface manages to retrieve (GET) and set (PUT or TOGGLE) the values of those sub-resources. 
The management of multiple sub-resources requires SenML for this interface and as extension of the Link List interface. 
Third, the Linked Batch is an extension of the Batch interface which is dynamically controlled by the web client and has no sub-resources. 
The resources forming the batch are referenced using CoRE Link Format and RFC5988. 
This is contrary to the basic Batch that is a static collection defined by the web server. 
Fourth, the Sensor interface retrieves values from a sensor device. Either plain text and SenML formats can be defined as the Media Type but in order to retrieve single measurements requiring no meta-data, the use of plain text is recommended. 
Fifth, the Parameter interface manages configurable parameters and other information where each parameter can be read (GET) or set (PUT). 
Sixth, the Read-Only Parameter interface is conceptualized for parameters that can be read (GET) but not set (PUT). 
Seventh, the Actuator interface models different kinds of actuators where the change of a value has an effect on its environment. 
Several actuators (e.g. LEDs, relays, light dimmers, motor controllers, etc) can be manipulated through the read (GET) and set (PUT) methods. 
Eighth, the Binding interface manipulates the binding table where each new binding is appended by a POST method and a content type of application/link-format. 
This requires that all the links contained in the payload must have relation type $boundTo$. 
The GET request returns the current status of a binding table, and the DELETE request removes the table.

In addition, Bormann~\cite{2b} proposed to represent collections of Link List in JSON format (RFC4627) for exchanging between resource directories.
JSON is considered by the Working Groups from the IETF such as the Constrained Resources (CoRE) Working Group, and the Constrained Management (COMA) Working Group as the most suitable protocol to structure the data exchange leaving other formats such as XML optional.
Unlike IoT constrained devices, global and local directories are able to manage more information and bandwidth.
In these cases, the usage of JSON to represent the Link List information is more useful. 
Bormann defines a simple mapping in JSON that contains the information of the formats specified in WebLinking and CoRE Link Format.
Mapping each web link (“link-value”) is a collection of attributes (“link-param”) applied to a “URI-Reference”.
In other words, a JSON Object formed by name/value pairs (member) where the parameter name or attribute is named “parname”, the value of the parameter or attribute value is named “ptoken” or “quoted-string”. 
This last option can cause that the results need to be parsed as defined in CoRE Link Format. 
When an attribute is duplicated, its values are represented as a JSON array of string values. 
The URI is represented by the pair name/value “href” and the URI-Reference. 
Table~\ref{tab23} illustrates the JSON mapping of Link List collections.

\begin{table*}[htbp]
\begin{displaymath}
\begin{array}{|l|}
\hline
\textbf{Parse Example} \\
\hline
</sensors>;ct=40;title="Sensor Index", \\
</sensors/temp>;rt="temperature-c";if="sensor", \\
</sensors/light>;rt="light-lux";if="sensor", \\
<http://www.example.com/sensors/t123>;anchor="/sensors/temp";rel="describedby", \\
</t>;anchor="/sensors/temp";rel=”alternate” \\
\hline
"[{"href":"/sensors","ct":"40","title":"Sensor Index"}, \\
  {"href":"/sensors/temp","rt":"temperature-c","if":"sensor"}, \\
  {"href":"/sensors/light","rt":"light-lux","if":"sensor"}, \\
  {"href":"http://www.example.com/sensors/t123","anchor":"/sensors/temp","rel":"describedby"}, \\
  {"href":"/t","anchor":"/sensors/temp","rel":"alternate"}] " \\
\hline
\end{array}
\end{displaymath}
\caption{JSON mapping for Link Lists} 
\label{tab23}
\end{table*}

\subsubsection{Semantic Web of Things}
Semantic Web of Things was proposed under the framework of an European project called SPITFIRE~\cite{63d}. 
SPITFIRE aims to integrate the current Internet with the embedded computing world. 
The project proposed a discovery mechanism for sensors and things based on the definition of a specific ontology. 
This ontology provides a high abstraction for integrating heterogeneous sensors and things into the Linked Open Data (LOD) cloud which is an effort to link semantic data on the web.
Specifically, the ontology is based on Resource Description Framework (RDF) to enhance the interaction with sensors through the web.
RDF is the main technique for machine-readable representations of knowledge on the web. 
In particular, RDF represents knowledge as triples (subject, predicate, object).
A set of triples forms a graph where subjects and objects are vertices and predicates are edges. 
From the graph formed by these triples, one can infer information by exploiting the knowledge that is a transitive property from a RDF graph.
The graph is imperative to use non-ambiguous identifiers for subjects, predicates and objects to guarantee uniqueness on an Internet scale.
This is achieved by encoding triples as URIs. 
An example of triples could be expressed as follows:
\begin{itemize}
  \item A Subject: "http://example.com/sensors/sensor3"
  \item A Predicate: "http://example.com/locations/hasLocation21"
  \item An Object: "http://example.com/parkingSpot/spot41"
\end{itemize}

Through queries by SPARQL, the triples server obtains information related to the sensor. 
Assuming that sensors are described by such RDF triples, a search service and find sensor are based on meta-data such as sensor type, location or accuracy. Queries can be expressed in SPARQL (similar to SQL) and provides a powerful way to search knowledge between RDF triples.
Table~\ref{tab24} presets a SPARQL query for RDF triples that stores the subjects observing the occupancy of parking places in Berlin. 
The simple SPARQL query allows to find the free spots near a certain location.

\begin{table*}[htbp]
\begin{displaymath}
\begin{array}{|l|}
\hline
SELECT\ COUNT(DISTINCT\ ?node)\ as\ ?spots \\
WHERE\ \{ \\
\ \ \ \ \ \ ?node\ a\ ssn:Sensor\ ;  \\
\ \ \ \ \ \ \ \ ssn:observes\ ex:Occupancy\ ; \\
\ \ \ \ \ \ \ \ dul:hasLocation\ ?spot\ . \\
\ \ \ \ \ \ ?spot\ a\ ex:ParkingSpot\ ; \\
\ \ \ \ \ \ \ \ dul:hasLocation\ dbpedia:Berlin\ . \\
\} \\
\hline
\end{array}
\end{displaymath}
\caption{Example of searching by SPARQL in RDF triples.} 
\label{tab24}
\end{table*}

\subsubsection{EXI: Efficient XMl Interchange}
The Efficient XML Interchange (EXI)~\cite{3b} format is a very compact, high performance XML binary representation.
EXI reduces significantly bandwidth requirements without compromising efficient use of other resources such as code size, battery life, processing power and memory. 
Moreover, EXI uses a grammar-driven approach that achieves very efficient encodings that utilize a straightforward encoding algorithm and a small set of datatype representations. 
EXI allows using available schema information to improve compactness and performance.
Also, EXI can be re-used current digital signature techniques from XML to reduce the development effort. 
In particular, EXI is employed in some IoT projects (i.e. IoT@Work) to represent the capabilities from a sensor to access to another one.

\subsubsection{oBIX: Open Building Information Xchange}
The Open Building Information Xchange (oBIX)~\cite{26p} is a specification published by the Organization for the Advancement of Structured Information Standards (OASIS) in December 2006. 
This platform-independent technology is designed to provide M2M communications between embedded software systems over existing networks using standard technologies such as XML and HTTP.
oBIX is based on service-oriented client/server architecture and defines only three request/response services used to read and manipulate data or to invoke operations. 
Each service response is an oBIX XML document that contains the requested information or the result of the service. 
The implementation of these three request/response services is called binding.
There are two different bindings specified by the oBiX standard. 
The first HTTP binding maps oBIX request to HTTP methods. 
The second SOAP binding maps each oBIX request to a SOAP operation.

\begin{table*}[htbp]
\begin{displaymath}
\begin{array}{|l|}
\hline
<readReq> \\
\ \ \ <id>Mandat\_International\_Hall.Floor1.sensor1</id> \\
</readReq> \\
\hline
\end{array}
\end{displaymath}
\caption{Example of the oBIX use to read a sensor} 
\label{tab25}
\end{table*}

A fundamental element in the oBIX specification is the concise but extensible object model. 
These objects are described by attributes, called facets. 
Objects are identified by a name, a URL or both. 
Each object can contain other objects and the object model can be extended by a mechanism called contracts. 
The contract is utilized to define new types but provides a possibility of specifying default values.
The second essential part of oBIX specification is the simple XML syntax to represent the object model. 
Basically each oBIX object maps to exactly one XML element. 
Sub-objects result in the embedding of XML elements. 
Table~\ref{tab25} shows an example request to read a sensor in the first floor of Mandat International in Geneva.

\subsubsection{Comparative between Semantic Technologies for IoT Data Exchange}
\begin{table*}[htbp]
\begin{displaymath}
\begin{array}{|l|c|c|c|c|c|c|}
\hline
\textbf{Features}	& \textbf{IPSO}	& \textbf{IPSO} & \textbf{Weblinks} & \textbf{RDF} & \textbf{EXI} & \textbf{oBIX} \\
& \textbf{Text\ Plain}	& \textbf{SenML+JSON} & & & & \\
\hline
Format    & Text\ plain & RFC5988/JSON & RFC5988/JSON & XML & XML & XML \\
\hline
Wide\ extended & YES & YES & NO & NO & NO & YES \\
\hline
Parser\ embedded & NO & YES & YES & YES & YES & YES \\
\hline
Search\ Engine & CoAP\ RD & JSON\ based\ as & mDNS & SPARQL & XML & XML \\
               &         & ElasticSearch  & & & & \\
\hline
Communication & Low & Medium & Low & High & High & Very\ High \\
\hline
Semantic &	Very\ Low & High & High & Very\ High & Very\ High & Very\ High \\
\hline
Memory & Low & Medium & Medium & Very\ High & High & Very\ High \\
\hline
\end{array}
\end{displaymath}
\caption{Data exchange technologies for IoT semantic description} 
\label{tab26}
\end{table*}

This section shows a comparison between aforementioned data exchange technologies to provide homogeneous semantic in heterogeneous IoT domains, as shown in Table~\ref{tab26}.
In particular, oBIX is the most used in existing sensor networks due to its powerful description for Building Automation Systems. 
oBIX offers a relevant alternative to legacy technologies (i.e. BACnet, X10, KNX, etc) to achieve an open Building Automation through Web Services.
oBIX is based on HTTP and SOAP, therefore is highly interoperable and relevant for this work. 
The main problem is that oBIX is very heavy for constrained environments, and is based on SOAP and not available for CoAP. 
For constrained IoT devices, IPSO alliance is considered the most suitable for CoRE Working Group technologies and industrial sectors.
IPSO SenML+JSON is the most adequate solution considered since it has higher capabilities to describe the native semantic from IoT resources and services. 
Therefore, we consider the IPSO Alliance approaches for semantic description in the proposed oriented-service architecture.
In addition, JSON descriptions can be employed to design an search engine based on context awareness, as described next Section.

\subsection{Search Engine based on ElasticSearch}
\label{sec:search}
This section presents the search engine used to support global queries in the proposed architecture. 
The architecture must provide fast and customized searches with context awareness in terms of geo-location, domain and type of resources. 
The search engine is based on ElasticSearch~\cite{4b}, a document oriented database that enables global queries through JSON language.
ElasticSearch is an open source search engine for distributed RESTFul-based architectures. 
The main features of ElasticSearch are:
\begin{itemize}
  \item Semantic description based on RESTful and JSON to format the queries and responses. 
  \item Ease configuration to minimize the launch of a search.
  \item Distributed solution to enable hundreds of nodes offering high availability, supporting large amounts of data.
  \item Real-time searches with short response times.
  \item Very versatile and sophisticated querying
  \item Easy management by a native API for Java.
  \item Geo-distance sorting
\end{itemize}

\subsubsection{ElasticSearch for Query DSL based on JSON}
ElasticSearch provides a full Query DSL based on JSON to define queries. 
The structure of JSON allows for high complex queries to filter and obtain the specific results in a low time.
Query DSL is a framework which enables the construction of type-safe SQL-like queries for multiple backends in Java.
In general, there are basic queries such as term or prefix. 
There are also compound queries like the $bool$ query. 
Queries can also have filters associated with them such as the filtered or constant score queries with specific filter queries.
Certain queries can contain other queries (like the $bool$ query) while others can contain filters (like the constant score), and some can contain both a query and a filter (like the filtered). 
Each request can contain any query from the list of queries or any filter from the list of filters.
ElasticSearch provides a high ability to build quite complex queries.

\subsubsection{ElasticSearch for Filters and Caching}
Filters can be an ideal candidate for caching. 
Caching the result of a filter require few memory, and may cause other queries executing against the same filter (same parameters) to be extremely fast.
Some filters already produce a result that is easily cacheable.
The decision of caching or non-caching is in the act of placing the result in the cache or not. 
These filters which include the terms, prefixes, and range filters are by default cached and are recommended to use when the same filter will be used across multiple different queries. 
For example, a range filter with age is higher than 10.
Other filters, usually already working with the field data loaded into memory are not cached by default. 
Those filters are already very fast, and the process of caching them requires extra processing in order to allow the filter result to be used with different queries than the one executed. 
These filters, including the geolocation, numeric range, and script filters, are not cached by default.
The last types of filters are those working with other filters. 
The "and", "not" and "or" filters are not cached as they basically just manipulate the internal filters.
ElasticSearch is able in an optimal time to get cacheable results filtered by resource type (e.g. light).

\subsubsection{ElacticSearch for Global Discovery}
The oriented-service architecture needs to provide a flexible discovery solution to carry out global organized look-ups( i.e. queries filtered by some attribute). 
For this purpose, the current solution for the Internet of Things is the CoAP Discovery described in the RFC6690~\cite{75d}. 
CoAP Discovery defines the look-up/query based on resource types. 
This allows filtering the resources to be discovered specifying a resource type (rt) in the query. 
However, CoAP Discovery is limited to discover in local domains by multicast messages.

Using ElasticSearch, the proposed architecture enables the global look-ups filtered by types and locations of resources and services. 
ElasticSearch supports quick queries and responses between the \textit{digcovery} and local \textit{digrectories}.
\textit{Digcovery} employs ElasticSearch to collect services information from the different \textit{digrectories} and make feasible its look-up and filtering based on the type and domain of services.
ElasticSearch is an interesting tool to store and retrieve stored data quickly.
ElasticSearch provides a fast database with multiple options of access, as well as a search engine based on RESTful.
The main advantage of ElasticSearch is that it offers the mechanisms required to manage a distributed and heterogeneous set of \textit{digrectories}, in an optimal time, and organizes results filtered by resource type (e.g. light). 
Thereby, this search engine supports global look-ups in ubiquitous and heterogeneous IoT domains.

\subsubsection{Geo-location in ElasticSearch for Context Awareness Discovery}

ElasticSearch offers a context awareness solution to discover resources and services based on their geo-locations.
In addition of the filtering by resource type, the geo-location allows discovering services that are close to you.
The geo-location of close resources is very useful for end-users in IoT environments such as smart cities.

The meaning of close is very different from the networking and physical point of view.
Since close in networking means under a common domain over a link-local which is usually mapped to a specific location.
But when you extend it through virtual networks and tunnels, this lost the meaning of close in terms of distance. 
At the same time, with the proliferation of Wireless networks such as 3G, LTE, Wi-Fi and Wimax, you can be located next to one device, but belongs to domains totally different.
For that reason, the geo-location solution requires a global service discovery that integrate multiple heterogeneous domains.
To do that, the global \textit{digcovery} integrates multiple domains from different \textit{digrectories} including resources and services with different locations and then apply the geo-location query of ElasticSearch based on the distance concept, as shown in Table~\ref{geolocation}.
The proposed architecture supports context awareness searches by the geo-location of heterogeneous IoT devices over latitude longitude coordinates.
This context awareness discovery is employed for an own client application called \textit{digcovery mobile} presented in the next section.

\begin{table*}[htbp]
\begin{displaymath}
\begin{array}{|l|}
\hline
query:\{ \\
\ \ filtered:\{ \\
\ \ \ \ query:\{ \\
\ \ \ \ \ \ range:\{ \\
\ \ \ \ \ \ \ \ longitude:\{ from:37.997 , to: 37.999 \}\} \\
\ \ \ \ \} , \\
\ \ \ \ filter:\{ \\
\ \ \ \ \ \ range:\{ \\
\ \ \ \ \ \ \ \ latitude:\{  from:−1.142 , to:−1.140 \}\} \\
\ \ \ \ \} \\
\ \ \} \\
\} \\
\hline
\end{array}
\end{displaymath}
\caption{Example of geo-location query} 
\label{geolocation}
\end{table*}

\subsection{Communications Interfaces for Integrating Heterogeneous Things and Clients}
\label{sec:communication-interfaces}

\begin{figure*}[htbp]
\begin{center}
\includegraphics[width=1\textwidth]{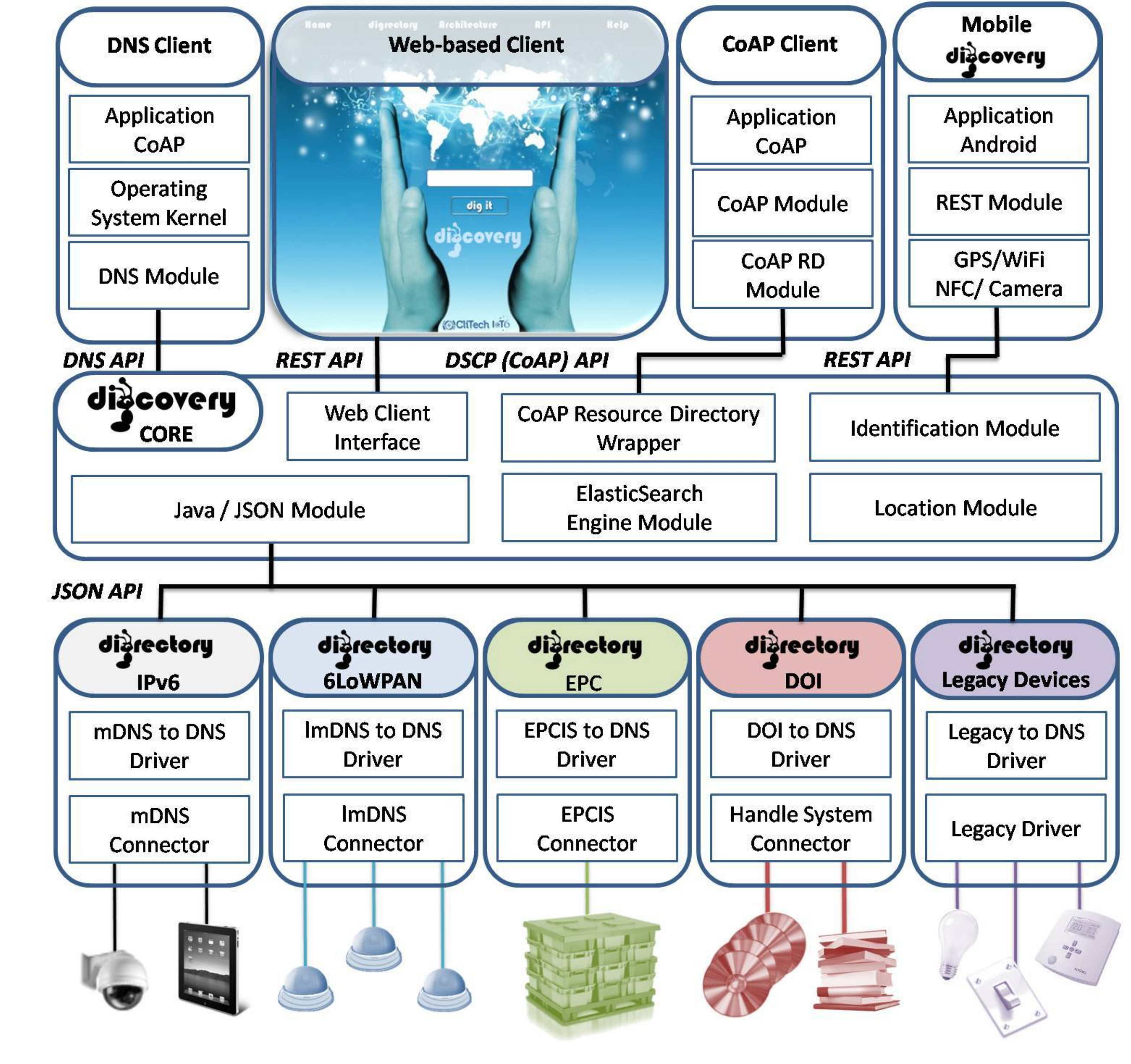}  
\caption{Communication interfaces between the IoT components} 
\label{fig16:General-APIs}
\end{center}
\end{figure*}

This section presents the communication interaction between the IoT components of the proposed architecture shown in Figure~\ref{fig16:General-APIs}.
The proposed architecture offers standardized discovery and registry services (i.e. DNS) for heterogeneous IoT devices.
The different technologies involved in the Internet of Things ecosystem such as Smart Objects, RFID tags, and legacy devices are integrated through \textit{digrectories}. 
These \textit{digrectories} are managed through DNS-queries extended with an ElasticSearch engine to provide a scalable architecture at the same time that this enables a centralized point, called \textit{digcovery} core, to manage and discover them.

\textit{Digrectories} are the components deployed locally in each domain to handle the resources in subnet level.
These \textit{digrectories} provide specific drivers to connect with things located in the domain.
The \textit{digrectories} allow the communication with services under heterogeneous communication technologies such as the IPv6, 6LoWPAN,  EPCIS or legacy (e.i. CAN, X10, EIB/KNX and BACNet).
\textit{Digrectories} manage each technology through a specific protocol: IPv6 devices through mDNS, 6LoWPAN devices through the proposed lightweight-mDNS, RFID tags through the EPCIS, and legacy devices through their proprietary protocols (i.e. CAN, X10, EIB/KNX and BACNet).
The \textit{digrectories} map the different services to a common DNS structure through specify drivers according to the IoT technologies. 
These drivers translate from the original protocol to an unified DNS-based protocol to provide an homogeneous face to the \textit{digcovery core}.
DNS structures are encapsulated in JSON objects to optimize the communication process.
A JSON interface is used for the intra-communication between the \textit{digrectories} and the \textit{digcovery} to inform about the resources accessible in local domains.

\textit{Digcovery} is a central system which can be extended in the cloud to manage the different domains and subnets.
This can be seen as a searching platform such as Google and Yahoo, but \textit{digcovery} is oriented to IoT services and resources. 
\textit{Digcovery} enables to discover the proper functionalities of IoT devices before clients interoperates directly.
\textit{Digcovery} is not the same as a M2M platform which acts as a proxy for the communication between the things and the clients.
Once clients found the IoT things through \textit{digcovery}, clients can communicate directly with smart things without going through the \textit{digcovery} core. 
The communication between clients and things is based on a COAP interface~\cite{20p} developed by the IETF CoRE Working Group. 
This enables the integration of constrained devices with an overhead of only 4 bytes and a functionality optimized for the observation of resources ~\cite{21p}, application-layer fragmentation ~\cite{22p}, and mapping with the HTTP-based RESTful architecture.

The proposed architecture has been developed to enable global look-ups and queries based on a common semantic description.
In this architecture, all the resources and services are mapped to a common semantic description based on the IPSO SenML+JSON approach~\cite{17p, 18p}.
These semantic description has been integrated into the DNS-SD types to reach a common semantic description accessible through DNS and powered with the universal IPv6 capabilities to carry out the discovery resolution.
\textit{Digcovery} exploits the ElasticSearch engine to provide global queries on various \textit{digrectories} with context awareness based on the type and location of services. 
This proposed architecture allows global context-awareness queries over heterogeneous and distributed IoT resources mapped to a common semantic.

In the proposed architecture, the client applications are highly heterogeneous depending on the technology used. 
The first top module presents the usage of DNS technology to exploit existing IP-based protocols and mechanisms. DNS is the main discovery technology used by IPv6-enabled clients based on DNS-SD format and mDNS messages.
The second top module is the Web-based platforms to access and register resources through the RESTFul architecture. 
The Web interface is offered for the majority of HTTP clients.
For constrained clients, CoAP is supported since it is the main technology for discovery from the CoRE working group. 
Regarding the COAP integration, this also offers the \textit{Digcovery} CoAP Service Protocol (DCSP) for CoAP-enabled devices.
Therefore, CoAP Resource Directory are wrapped around the DNS functions through the presented wrapper in the figure.
The final module supported by the proposed architecture is the \textit{mobile digcovery}.
The \textit{mobile digcovery} extends the architecture with the identification and location capabilities. 
These capabilities are offered by the mobile platforms such as smart phones through the integrated technologies: GPS and WiFi (real time location systems) for location, and RFID and cameras (barcodes and QR codes) for identification.

\begin{figure*}[htbp]
\begin{center}
\includegraphics[width=0.7\textwidth]{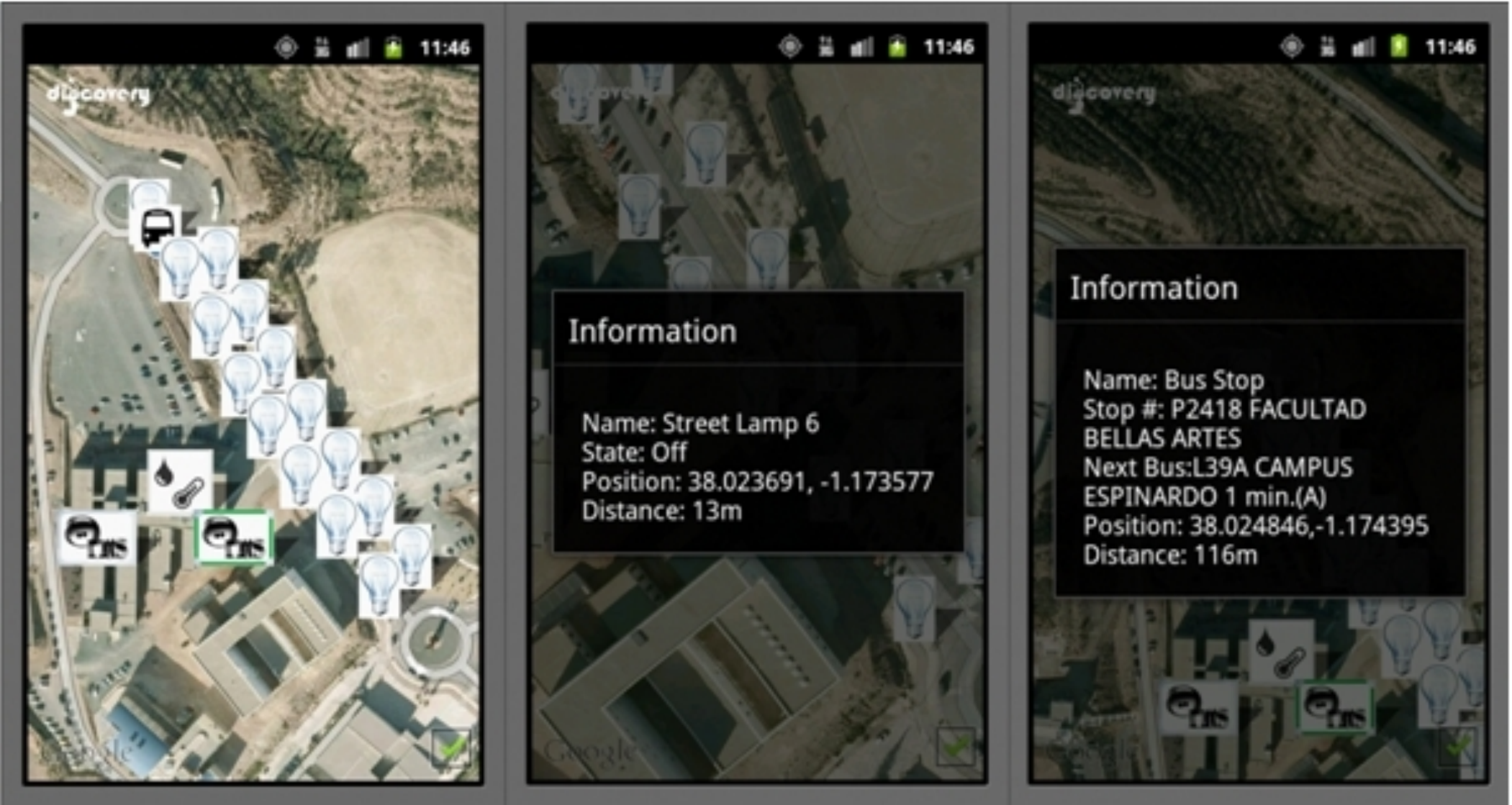}  
\caption{Example of the mobile application}
\label{mobile}
\end{center}
\end{figure*}

The \textit{mobile digcovery} application is currently under development.
This mobile application allows the discovery of heterogeneous resources and services through the \textit{digcovery} core to obtain information and interoperate with the IoT devices (i.e. turn off/on a light).
Figure~\ref{mobile} presents the mobile application discovering smart things such as street lights and bus stops.
These smart things are deployed at the street next to the Computer Science Faculty from the University of Murcia. 
The \textit{mobile digcovery} is able to make context-awareness searhes by geo-location and also access directly to smart things through CoAP interface. 

\subsection{Satisfaction of IoT Requirements}

This section enumerates how the proposed architecture satisfies the requirements of heterogeneous and ubiquitous IoT networks.

\begin{itemize}
  \item Scalability. The proposed architecture is based on the decentralized DNS infrastructure. 
  The distributed solution allows a scalable discovery of IoT resources and services through multicast DNS(mDNS) at the device level and through the hierarchical delegation of DNS Service Discovery (DNS-SD) in the \textit{digrectories} and \textit{digcovery} at the local and global level, respectively.
Where, \textit{digrectories} can be located at gateways or routes, and the \textit{digcovery} system may be located at a cloud platform as shown in Section~\ref{sec:digcovery-digrectory}.

  \item Dynamic. mDNS allows updating the changes of IoT resources and services between IoT devices and \textit{digrectories}.
Moreover, the proposed architecture based on DNS-SD mechanism enables the easy updates of available services in the \textit{digrectories} and \textit{digcovery} as explained in Section~\ref{sec:digcovery-mechanisms}. 

  \item Communication constraints. The original frame size from IoT technologies such as IEEE 802.15.4 is 127 bytes, and networking header reduces from 61 to 76 bytes of payload.
  The DNS protocol usually includes sections with additional and authority records with high overload. 
  To reduce the overload, we recommend to exclude these records in the DNS messages. 
  Moreover, we provide several optimizations of mDNS and DNS-SD by reducing the size of SRV and TXT entries.
  In particular, our optimizations enable the sending of mDNS and DNS-SD messages in an unique 802.15.4 frame of 127 bytes, while the original use of DNS-SD and mDNS needs 3 or more frames, as shown in Section~\ref{sec:smart-objects-discovery-protocol}.

  \item Semantic description. A semantic based on the IPSO SenML+JSON approach is chosen to describe the services and attributes to carry out homogeneous queries, as presented in Section~\ref{sec:semantic}. 
  
  \item Global query capacity. 
  The proposed architecture supports global queries by the \textit{digcovery} platform using the ElasticSearch engine through the different \textit{digrectories}. ElasticSearch enables context awareness queries based on geo-location and resource types.
  Further details about the ElasticSearch approach for supporting global queries are described in Section~\ref{sec:search}. 

  \item Heterogeneity of things and clients. \textit{Digrectories} support specific protocols to communicate with heterogeneous IoT devices based on different technologies such as IPv6, 6LoWPAN,  EPCIS or legacy (e.i. CAN, X10, EIB/KNX and BACNet). The \textit{digcovery} core offers the most used discovery interfaces (i.e. DNS, HTTP and COAP) to support the interoperability with the majority of Internet clients, as described in Section~\ref{sec:communication-interfaces}.

  \item Based on existing technologies. The proposed architecture is based on IP-based technologies and protocols such as DNS and Web Services. 
  In particular, the architecture provides the DNS extensions: mDNS and DNS-SD. 
  Regarding to the Web Services, the architecture supports HTTP/RESTful solutions for Internet hosts and COAP solutions for constrained devices, as described in Section~\ref{sec:communication-interfaces}.

\end{itemize}

\section{Conclusions}
\label{conclusions}

The paper presents a scalable oriented-service architecture for discovering, registering and looking-up services and resources of heterogeneous and ubiquious IoT devices.
For the architecture, we propose three main elements: global \textit{digcovery}, local \textit{digrectory} and smart object discovery.
Global \textit{digcovery} is a centralized cloud-platform which distributes hierarchically the looking-up to the \textit{digrectories} following the scalable DNS infrastructure.
In a local domain, each \textit{digrectory} registers fine-grained descriptions of the resources and services based on DNS-SD format. 
The smart object protocol enables the discovery and access of resources and services available from IoT devices using the lightweight mDNS protocol.

In addition, we analyse and provide other necessary functionalities such as semantic description, context-awareness search and communication interfaces to achieve an unifying architecture.
First, we compare existing semantic technologies for data exchange and provide the IPSO SenML+JSON approach which is the most suitable semantic for heterogeneous IoT domains according to the CoRE Working Group.
Second, we analyse the ElasticSearch protocol and its integration in the proposed architecture to support context awareness look-ups based on geo-location, domain and type of resources.
Third, we provide the communication interfaces to enables the interoperability between the proposed elements with heterogeneous IoT things and clients. 

The proposed architecture is compatible with existing protocols based on standardized technologies such as IPv6 and DNS.
Moreover, the architecture supports the integration of heterogeneous IoT devices including 802.15.4 sensors, RFID tags, building actuators, and mobile phones. 
The architecture also provides an open service layer to interact with end-user applications through standardized interfaces such as web services (HTTP), and constrained applications (COAP).

For future work, we consider security and privacy which are horizontal challenges in open and ubiquitous IoT services. 
A secured architecture should manage the access control to the IoT resources and services such as energy management sensors at smart grids, smart patient’s monitors at hospitals and traffics sensors at transportation systems.
Without control access, malicious operations could cause critical problems(i.e infrastructure damage and life loss).
Therefore, we will study existing solutions for access control and the integration in the proposed architecture to mitigate the risks of unauthorized access to IoT resources and services.

\section{Acknowledgments}

This work has been sponsored by the Excellence Researching Group Program (04552/GERM/06) from Foundation Seneca, and with funds from the IoT6 European project (STREP) from the 7th Framework Program (Grant 288445).







\bibliographystyle{elsarticle-num}








\end{document}